\documentstyle[12pt]{article}
\newcommand{\be}{\begin{equation}}
\newcommand{\ee}{\end{equation}}

{
\begin{document}

\begin{center}

{\large\bf Exact quantum states of a general time-dependent 
quadratic system from classical action}
\end{center}
\begin{center}
Dae-Yup Song\footnote[2]{Electronic address: 
        dsong@sunchon.sunchon.ac.kr}
\end{center}

\begin{center}
{\it Department of Physics,\\ Sunchon National University, Sunchon 
540-742, Korea}
\end{center}

\begin{abstract}
A generalization of driven harmonic oscillator with time-dependent 
mass and frequency, by adding total time-derivative terms to the 
Lagrangian,  is considered. The generalization which gives a general 
quadratic Hamiltonian system does not change  the classical 
equation of motion. Based on the observation by
Feynman and Hibbs, the propagators (kernels) of the systems are 
calculated from the classical action, in terms of solutions
of the classical equation of motion: two homogeneous and one 
particular solutions. The kernels are then used to find wave 
functions which satisfy the Schr\"{o}dinger equation. One of the wave
functions is shown to be that of a Gaussian pure state. In every case 
considered, we prove that the kernel does not depend on the way of
choosing the classical solutions, while the wave functions depend 
on the choice. The generalization which 
gives a rather complicated quadratic Hamiltonian is simply 
interpreted as acting an unitary transformation to the driven 
harmonic oscillator system in the Hamiltonian 
formulation.
\newline
PACS number(s): 03.65.Ca, 03.65.Bz, 03.65.Db, 03.65.Ge

\end{abstract}
\newpage

\section{Introduction}

Time-dependent quantum mechanical systems continue of great interest. 
In particular, the system described by an explicit time-dependent 
quadratic Hamiltonian has attracted considerable attention. One of the typical 
examples is the harmonic oscillator with time-dependent mass and/or 
frequencies. Those studies  have many applications such as in quantum 
optics \cite{AK}, Paul trap \cite{Brown}, and the analyses of quantum 
fields in curved spacetime \cite{Fulling}, and they are closely related to 
the theory of quantum dissipation \cite{JCP}. 

In the Hamiltonian formulation of a time-dependent harmonic oscillator, 
Lewis and Riesenfeld (LR) \cite{LR,Lewis} have shown that there exists 
quantum mechanically invariant operator whose exact form is determined 
by an auxiliary function. The invariant operator can then be used to 
find exact wave functions of quantum states. During the past several years, 
this LR method has been widely used for the study of a general quadratic 
hamiltonian systems 
\cite{Ped, Yeon1,Ji,Lee,Yeon2}. 
For the harmonic oscillator with time-dependent mass and frequency, 
the wave functions and the kernel (propagator) have been found 
\cite{Ped,Yeon1,Yeon2}.
Through Heisenberg-picture approach, Ji {\em et al.} \cite{Ji} have refined 
the derivation, and the wave functions are given. The Heisenberg-picture 
approach have then been used \cite{Lee}, with successive 
unitary transformations \cite{Brown}, to find exact wave functions of 
the general quadratic Hamiltonian system \cite{Yeon2}.
The auxiliary functions in the LR method were defined
through differential equations related to the equation of motion.
We also note that the LR method can be applied for more general 
systems \cite{Lew}.

In this article, we will study the quadratic system 
in the Lagrangian formulation of Feynman and Hibbs \cite{FH}.
The Lagrangian we will consider is
\be
L= {1\over 2} M(t) \dot{x}^2 - {1\over 2}M(t)w^2(t) x^2 +F(t) x 
    + {d \over dt} (M(t)a(t) x^2) + {d \over dt} (b(t) x) +f(t),
\ee
where $w^2(t),F(t), a(t),b(t),f(t)$ are arbitrary real function of $t$ and 
$M(t)$ is also arbitrary real but always positive.
The last three terms in the right hand side (r.h.s.) of Eq.(1) have no 
effect in classical dynamics and classical trajectory $\bar{x}$ of the 
coordinate $x$ will satisfy the equation:
\be
{d \over {dt}} (M \dot{\bar{x}}) + M(t) w^2(t) \bar{x} =F(t).
\ee
The most general solution of this differential equation may be 
composed of a particular solution and two linearly independent
homogeneous solutions.
The corresponding {\em quantum} Hamiltonian may be written as 
\be
H= {\hat{p}^2 \over 2 M(t)} - a(t)[\hat{p}\hat{x}+\hat{x}\hat{p}]
     +{1\over 2} M(t)c(t)\hat{x}^2 -{b(t)\over M(t)} \hat{p}+d(t)\hat{x}
      +( {b^2(t) \over 2M(t)} -f(t)),
\ee 
where
\begin{eqnarray}
c(t)&=&w^2 + 4a^2 -2 \dot{a} -2 {\dot{M}\over M}a, \nonumber\\
d(t)&=& 2ab -\dot{b} -F.
\end{eqnarray}
In the sense of differentiation of operator in Ref.\cite{Landau}, the 
quantum equation of motion for the operator $\hat{x}$ is again given as 
\[
{d \over {dt}} (M \dot{\hat{x}}) + M(t) w^2(t) \hat{x} =F(t).
\]
Feynman and Hibbs have shown that the coordinate dependent part of 
the kernel is determined from the classical action (Ch.3-5 of 
Ref.\cite{FH}). In this 
article, it will be shown that the remained part of the kernel for the 
system will be completely
determined from the Schr\"{o}dinger equation and the initial condition 
which the kernel should satisfy. In this way the kernel for the 
system of Eq.(1) will evaluated. By the method of Ref.\cite{KL}, 
the wave functions will then be evaluated from the kernel. 
This Lagrangian formulation has a clear advantage over the Hamiltonian
formulation in showing how the last three terms in the r.h.s. of Eq.(1)
which have no effect on classical dynamics affect the wave functions.
The classical action will be evaluated in terms of the two 
linearly independent homogeneous 
solutions and one particular solution, and so are the kernel and the 
wave functions. We will prove that the kernel do not depend on the way 
of choosing the classical solutions, while the wave functions are 
{\em not} unique and depend on the choice of the classical solutions. 
By comparing with the results on the Gaussian
pure states of Ref.\cite{Sch}, it is suggested that choosing different 
classical solutions might amount to acting unitary transformations to 
the annihilation operator.

In the next section, we will consider the harmonic oscillator with
time-dependent mass and frequency, mainly to expose our method. 
It will also be shown that the kernel does not depend on the choice 
of two homogeneous solutions, whilst the wave functions depends 
on the choice.  In Sec. III, the driven 
harmonic oscillator will be considered.
In Sec. IV, the system of a general quadratic Lagrangian in Eq.(1)
will be considered and some previous errors will be corrected.
The general system will be shown equivalent to the driven harmonic
oscillator through an unitary transformation.
Sec. V will be devoted to a summary and discussions. We add an 
appendix to explain how to determine the time dependent part of 
kernel from
the Schr\"{o}dinger equation and the initial condition.

\section{The harmonic oscillator with time-dependent mass and frequency}
In this section, we will apply our method to the harmonic oscillator 
without driving force. For this model the Lagrangian is written as;
\be
L^S=  {1 \over 2} M(t) \dot{x}^2 - {1 \over 2} M(t) w^2(t) x^2.
\ee
The action (integral) from time $t_a$ to time $t_b$ is written as;
\be
S = \int_{t_a}^{t_b} L dt
\ee
which gives the equation of motion for the classical trajectory of the
model considered
\be
{d \over {dt}} (M \dot{\bar{x}}) + M(t) w^2(t) \bar{x} =0.
\ee
To find a simple expression for the classical action that is the integral 
along classical trajectory, we can rewrite the action as
\be
S^S= {1 \over 2} M \dot{x} x \mid_{t_a}^{t_b} 
   - {1 \over 2} \int_{t_a}^{t_b} x[ {d  \over dt}(M\dot{x}) + Mw^2 x].
\ee
The classical action is then simply given as
\be
S_{cl}^S (a,b)= {1 \over 2}  M(t_b)x_b \dot{\bar{x}}_b 
             - {1 \over 2}  M(t_a)x_a \dot{\bar{x}}_a,
\ee
where $x_a$ ($x_b$) and $\dot{\bar{x}}_a$ ( $\dot{\bar{x}}_b$) denote 
fixed end point and $d \bar{x} \over dt$ at $t=t_a$ ($t=t_b$), respectively.

Suppose that $u(t)$ and $v(t)$ are two linearly independent real solutions of 
Eq.(7), so that $\bar{x}$ can be written by a linear combination of them.
From the two linearly independent solutions, one can always find $v_s(t)$, 
a linear combination of the two solutions which satisfies $v_s(t_a)=0$, so that 
$u(t)$ and $v_s(t)$ are again two linearly independent solutions.
One can then easily find the quantities $\Omega_s$ and $\Omega$
defined as 
\be
\Omega_s = M(t)[ \dot{v}_s(t)u(t) - \dot{u}(t)v_s(t)],~~~~
\Omega = M(t)[ \dot{v}(t)u(t) - \dot{u}(t)v(t)]
\ee
do not depend on time.

The $\bar{x}(t)$ with two fixed end points $x_a$, $x_b$ can be written as
\be
\bar{x}(t)= x_a {u(t) \over u(t_a)} 
   + [x_b- x_a {u(t_b) \over u(t_a)}]{v_s(t) \over v_s(t_b)}
\ee
Making use of this expression of $\bar{x}$, one can rewrite the classical 
action in Eq.(9) as 
\begin{eqnarray}
S_{cl}^S(a,b)
     &=& {x_a^2 \over 2} M(t_a) [- {\dot{u}(t_a) \over u(t_a)} 
                     +{u(t_b) \over u(t_a)}{\dot{v}_s(t_a) \over v_s(t_b)}]           
        +{x_b^2 \over 2}M(t_b){\dot{v}_s(t_b) \over v_s(t_b)}
            \nonumber \\
      &&+{x_a x_b \over 2}[M(t_b)({\dot{u}(t_b) \over u(t_a)} 
                 -{u(t_b)\dot{v}_s(t_b) \over u(t_a) v_s(t_b)} )
             -M(t_a){\dot{v}_s(t_a) \over v_s(t_b)}].
\end{eqnarray}     
The fact that classical dynamics are deterministic implies 
$S_{cl}$ is unique, as can be explicitly proved;
First, one find that $S_{cl}^S$ does not depend on the scaling of $u(t)$ 
or $v_s(t)$ by multiplying constant factors. Second, the classical action in 
Eq.(12) is invariant under the substitution of $u(t)$ by $u(t) +C v_s(t)$ 
with an arbitrary constant $C$. 
These two observations lead us to a conclusion that $S_{cl}^S$ do not depend 
on the particular choice of $u(t)$ or $v_s(t)$ as long as  $v_s(t_a)=0$.

Then from the formula (3-51) of Ref.\cite{FH}, the kernel can be written as;
\be
K^S(b,a)= \exp [ {i \over \hbar}(S_{cl}^S(a,b) +D^S(t_a,t_b))],
\ee
where, as shown in appendix, $D(t_a,t_b)$ can be completely determined from
the initial condition and Schr\"{o}dinger equation. Making use of the
formulas in appendix,
one can find the expression of $K^S(b,a)$ in terms of $u$ and $v_s$ as
\begin{eqnarray}
K^S(b,a)&=&   
       \sqrt{{M(t_a) \over 2\pi i \hbar}{\dot{v}_s(t_a) \over v_s(t_b)}}
        \exp[{i \over 2\hbar}
              ( M(t_a) (- {\dot{u}(t_a) \over u(t_a)} 
                  +{u(t_b)\dot{v}_s(t_a) \over u(t_a) v_s(t_b)} )x_a^2  
         \nonumber\\         
      &&~~~~~~~~~~~~~~~~~~~~~~~~~~~~~~
           +M(t_b){\dot{v}_s(t_b) \over v_s(t_b)}x_b^2 
                     -2x_ax_bM(t_a){\dot{v}_s(t_a) \over v_s(t_b)})].
\end{eqnarray}
Since the kernel is uniquely determined from the classical action, our 
argument on the uniqueness of the classical action implies that $K^S(b,a)$ 
does not depend on the way of choosing $u(t)$ or $v_s(t)$.

To find the wave functions from the kernel along the method of Ref.\cite{KL},
we define two functions $\rho(t)$ and $z(t)$ as;
\begin{eqnarray}
\rho(t)&=&\sqrt{u^2(t) +v_s^2 (t)}, \\
z(t)&=& {u(t) -iv_s(t) \over \rho(t) }.
\end{eqnarray}
After a little algebra, one can find that the kernel can be written as
\begin{eqnarray}
K^S(b,a)
&=& {1 \over \sqrt{\pi \hbar}}\sqrt{{\Omega_s \over \rho(t_a)\rho(t_b)}}
   \nonumber\\
& &\times\exp[{x_a^2 \over 2 \hbar}(-{\Omega_s \over \rho^2(t_a)}
                       -i M(t_a) {\dot{\rho}(t_a) \over \rho(t_a)})
           +{x_b^2 \over 2 \hbar}(-{\Omega_s \over \rho^2(t_b)}
                  +i M(t_b){\dot{\rho}(t_b) \over \rho(t_b)})]  \nonumber\\
& &\times\sum_{n=0}^\infty { z^{n+{1\over 2}}(t_b) \over 2^n n!}
                 H_n(\sqrt{\Omega_s \over \hbar} {x_a \over \rho(t_a)})
                 H_n(\sqrt{\Omega_s \over \hbar} {x_b \over \rho(t_b)}),
\end{eqnarray}
where $H_n$ is the n-th order Hermite polynomial.
From now on the definition of $\rho(t)$ is modified as 
$\rho(t)=\sqrt{u^2(t) +v^2 (t)}$.
From the well-known fact that
\be
K(x_b,t_b;x_a,t_a)=\sum_n \psi_n(x_b,t_b)\psi_n^*(x_a,t_a), 
         ~~~{\rm for}~~~ t_b>t_a,
\ee
one can find the n-th order wave function:
\be
\psi_n^S(x,t)= 
     {1\over \sqrt{2^n n!}}({\Omega \over \pi\hbar})^{1\over 4}
     {1\over \sqrt{\rho(t)}}[{u(t)-iv(t) \over \rho(t)}]^{n+{1\over 2}}
     e^{{x^2\over 2\hbar}[-{\Omega \over \rho^2(t)}
               +i M(t){\dot{\rho}(t) \over \rho(t)}]}
          H_n(\sqrt{\Omega \over \hbar} {x \over \rho(t)})
\ee
which satisfies the Schr\"{o}dinger equation:
\be
i\hbar {\partial \psi_n^S \over \partial t}= 
       -{\hbar^2 \over 2M(t)}{\partial^2 \psi_n^S \over \partial x^2}
       +{M(t) w^2(t)\over 2}x^2 \psi_n^S  .
\ee
The $\psi_n^S$ {\em does} depend on the choice of two homogeneous 
solutions, and any set of two linearly independent solutions can be 
used to construct the wave functions which satisfy the 
Schr\"{o}dinger equation of Eq.(20).

To have a physical interpretation of the fact that different choice of 
$\{u,v\}$ may give different set of wave functions 
$\{\psi_n^S,~~n=0,1,2,\cdots\}$, we consider the simplest case: 
the simple harmonic oscillator where $M(t)=m_0$
and $w(t)=w_0$. In this case, if we take $\{u,v\}$ as 
$\{C\cos w_0t, C\sin w_0t\}$ with arbitrary non-zero constant $C$, then the 
$\psi_n^S$ reduces to the usual stationary wave functions whose ground 
$(n=0)$ state is given as 
$\tilde{\psi_0}=({m_0w_0 \over \hbar \pi})^{1/4}e^{-{m_0w_0x^2 \over 2\hbar}}$. 
The choice of $\{u,v\}$ as $\{\cos w_0t, C\sin w_0t\}$ with 
$C\neq 1$, however, gives the wave functions of probability 
distribution pulsating as time passes. 

In general case, by defining $\gamma$ as $\gamma_1+i\gamma_2$ where 
\[\gamma_1= {\Omega\over \hbar\rho^2}~~{\rm and} ~~
\gamma_2=-{M\dot{\rho}\over \hbar \rho},\]
we can rewrite the $\psi_0^S$ as
\be
\psi_0^S = ({ \gamma_1 \over\pi})^{1/4}\exp(i\delta_0(t))
                    \exp[-{1\over 2} \gamma x^2],
\ee
with a real function $\delta_0$ of $t$. Therefore, $\psi_0^S$ is one of the 
wave functions of the Gaussian pure states extensively studied in Ref.\cite{Sch}.
There, it has been shown that any Gaussian pure state is the eigenstate of 
a certain linear combination of creation and annihilation  operator. 
If we choose different classical
solutions, then we could have different $\gamma$. 
The studies of Ref.\cite{Sch} suggest that choosing different classical solutions 
might amount to acting unitary transformations to the annihilation operator
of the representation system. 

\section{Driven harmonic oscillator}
In this section we will consider the system described by the Lagrangian:
\be
L^F= {1\over 2} M(t) \dot{x}^2 - {1\over 2}M(t)w^2(t) x^2 +F(t) x .
\ee
Let's denote the particular solution of Eq.(2) as $x_p(t)$, so that 
$x_p(t)$ satisfies the equation:
\[
{d \over {dt}} (M \dot{x_p}) + M(t) w^2(t) x_p=F(t).
\]
Then one can rewrite the Lagrangian as 
\begin{eqnarray}
L^F&=&{1\over 2} {d\over dt}[M(t)(x-x_p)(\dot{x}-\dot{x}_p)]
     +{d\over dt}[M(t)\dot{x}_p(x-x_p)] \nonumber\\
   & & -{1\over 2}(x-x_p)[{d \over {dt}} (M (\dot{x}-\dot{x}_p)) 
     + M w^2(x-x_p)]
     +{1\over 2}{d\over dt}[M\dot{x}_px_p]+{d\over dt}Y_{x_p}(t),
\end{eqnarray} 
where $Y(t)$ is defined as 
\be
Y_{x_p}(t)=\int_{t_0}^t{1\over 2} x_p(t')F(t')dt'
\ee
with arbitrary constant $t_0$.
The classical action $S_{cl}^F(a,b)$ from time $t_a$ to $t_b$ can 
be written as 
\begin{eqnarray}
\tilde{S}_{cl}^F(a,b;x_p)&=&
       S_{cl}^F(a,b)-\Delta S_1(x_p(t),Y_{x_p}(t))\mid_{t_a}^{t_b}
         \nonumber\\
     &=&[{1\over 2}M(t)(\bar{x}-x_p)(\dot{\bar{x}}-\dot{x}_p) 
              +M(t)\dot{x}_p(\bar{x}-x_p)]\mid_{t_a}^{t_b}.
\end{eqnarray}
By adding a homogeneous solution $Cu(t)+Dv(t)$ to the particular $x_p(t)$
with arbitrary constants $C$ and $D$, 
one can have a new particular solution $x_p' (t)$. By 
rewriting the classical action as 
\be
S_{cl}^F(a,b)=\int_{t_a}^{t_b}
       [{1\over 2}{d\over dt}(M\bar{x}\dot{\bar{x}}) 
         -{1\over 2}{d\over dt}(M\dot{\bar{x}}x_p)
         +{1\over 2}{d\over dt}(M\dot{x}_p\bar{x})
         +{1\over 2}x_pF]dt,
\ee
one can easily find that the classical action does {\em not} depend 
on the choice of particular solution. That is,
\be
S_{cl}^F(a,b)=\tilde{S}_{cl}^F(a,b;x_p)+
           \Delta S_1(x_p(t),Y_{x_p}(t))\mid_{t_a}^{t_b}
           =\tilde{S}_{cl}^F(a,b;x_p')
               +\Delta S_1(x_p'(t),Y_{x_p'}(t))\mid_{t_a}^{t_b}.
\ee
Through the same methods of previous section, one can find the end points 
dependent part of classical action:
\begin{eqnarray}
\tilde{S}_{cl}^F(a,b;x_p)
  &=&{M(t_a)[x_a-x_p(t_a)]^2 \over 2}[- {\dot{u}(t_a) \over u(t_a)} 
                  +{u(t_b)\dot{v}_s(t_a) \over u(t_a) v_s(t_b)} ]
 \cr
  & &+{M(t_b)[x_b-x_p(t_b)]^2 \over 2} {\dot{v}_s(t_b) \over v_s(t_b)}
 \cr
  & &-(x_a-x_p(t_a))(x_b-x_p(t_b)){M(t_a)\dot{v}_s(t_a) \over v_s(t_b)}
 \cr
  & &+M(t_b)\dot{x}_p(t_b)x_b-M(t_a)\dot{x}_p(t_a)x_a.
\end{eqnarray}
The kernel can be written as \cite{FH}
\be
K^F(a,b)= \exp [ {i \over \hbar}(S_{cl}^F(a,b) +D^F(t_a,t_b))]
      = \exp [ {i \over \hbar}(\tilde{S}_{cl}^F(a,b) 
                     +\tilde{D}^F(t_a,t_b))].
\ee
Since the $S_{cl}^F$ does not depend on the choice of the classical 
solutions within the given restriction and $D^F$ is uniquely determined 
from the $S_{cl}^F$, the kernel is again unique. For the explicit 
evaluation, we require the particular solution to satisfy $x_p(t_a)=0$.
In the notations of appendix, $B$ and $\beta$ is then given as;
\begin{eqnarray}
B&=&{M(t_b)\over 2}{\dot{v}_s(t_b) \over v_s(t_b)}, \\
\beta&=& -M(t_b)x_p(t_b){\dot{v}_s(t_b) \over v_s(t_b)} 
          -M(t_b)\dot{x}_p(t_b),
\end{eqnarray}
and the kernel is written as
\begin{eqnarray}
K^F(b,a)&=&   
       \sqrt{{M(t_a) \over 2\pi i \hbar}{\dot{v}_s(t_a) \over v_s(t_b)}}
  \nonumber\\
      & &\times\exp[{i \over 2\hbar}
                [  x_a^2 M(t_a)(- {\dot{u}(t_a) \over u(t_a)} 
                  +{u(t_b)\dot{v}_s(t_a) \over u(t_a) v_s(t_b)} )       
           +(x_b-x_p(t_b))^2M(t_b){\dot{v}_s(t_b) \over v_s(t_b)}
   \cr&&~~~~~~~~~~~~~~~~~ 
           -2x_a(x_b-x_p(t_b))M(t_a){\dot{v}_s(t_a) \over v_s(t_b)}
   \cr&&~~~~~~~~~~~~~~~~~~~
      +2M(t_b)\dot{x}_p(t_b)x_b -2M(t_a)\dot{x}_p(t_a)x_a 
        -M(t_b){\dot{v}_s(t_b) \over v_s(t_b)} x_p^2(t_b)
   \cr&&~~~~~~~~~~~~~~~~~~~
      -\int_{t_a}^{t_b}{M(t)\over v_s^2(t)}
        [x_p(t)\dot{v}_s(t)-\dot{x}_p(t)v_s(t)]^2 dt  ]].
\end{eqnarray}

From the expression of the kernel in Eq.(32), as in the
previous section, one can find the n-th order wave function as 
\begin{eqnarray}
\psi_n^F(x,t)&=& 
     {1\over \sqrt{2^n n!}}({\Omega \over \pi\hbar})^{1\over 4}
     {1\over \sqrt{\rho(t)}}[{u(t)-iv(t) \over \rho(t)}]^{n+{1\over 2}}
     \exp{[{(x-x_p(t))^2\over 2\hbar}(-{\Omega \over \rho^2(t)}
               +i M(t){\dot{\rho}(t) \over \rho(t)})]}
\cr
&&\times H_n(\sqrt{\Omega \over \hbar} {x -x_p(t) \over \rho(t)})
         \exp[{i\over\hbar}[M(t)\dot{x}_p(t)x 
         -{M(t)\over 2}{\dot{v}(t)\over v(t)} x_p^2(t)
\cr&&~~~~~~~~~~~~~~~~~~~~~~~~~~~~~~~~~~~~~~~
         -{1\over 2}\int_{t_0}^t M(z)(x_p(z){\dot{v}(z)\over v(z)} 
              -\dot{x}_p(z))^2dz]].
\end{eqnarray}
In Eq.(33), $\{u,v\}$ is the set of arbitrary linear independent 
homogeneous solutions, and $x_p$ is an arbitrary particular solution. 
The wave functions, again, depend on the way of choosing classical 
solutions and one can explicitly find that these wave 
functions indeed satisfy the Schr\"{o}dinger equation:
\be
i\hbar {\partial \psi_n^F \over \partial t}= 
       -{\hbar^2 \over 2M(t)}{\partial^2 \psi_n^F \over \partial x^2}
       +{M(t) w^2(t)\over 2}x^2 \psi_n^F -F(t)x\psi_n^F =H_F\psi_n^F.
\ee

\section{The general quadratic system}
In this section we will consider the general quadratic system 
described by the Lagrangian of Eq.(1). As in the previous sections, 
one can find the end points dependent part of classical action:
\begin{eqnarray}
\tilde{S}_{cl}^G(a,b;x_p)
  &=&{M(t_a)[x_a-x_p(t_a)]^2 \over 2}[- {\dot{u}(t_a) \over u(t_a)} 
                  +{u(t_b)\dot{v}_s(t_a) \over u(t_a) v_s(t_b)} ]
 \cr
  & &+{M(t_b)[x_b-x_p(t_b)]^2 \over 2} {\dot{v}_s(t_b) \over v_s(t_b)}
 \cr
  & &-(x_a-x_p(t_a))(x_b-x_p(t_b)){M(t_a)\dot{v}_s(t_a) \over v_s(t_b)}
 \cr
  & &+M(t_b)\dot{x}_p(t_b)x_b-M(t_a)\dot{x}_p(t_a)x_a
 \cr
  & &+M(t_b)a(t_b)x_b^2-M(t_a)a(t_a)x_a^2+b(t_b)x_b-b(t_a)x_a.
\end{eqnarray}
The only difference of $\tilde{S}_{cl}^G$ from $\tilde{S}_{cl}^F$ is the last 
four terms in the r.h.s. of Eq.(35).
Again, by requiring $x_p(t_a)=0$, one can evaluate the kernel in 
terms of classical solutions;
\begin{eqnarray}
K^G(b,a)&=&   
       \sqrt{{M(t_a) \over 2\pi i \hbar}{\dot{v}_s(t_a) \over v_s(t_b)}}
  \nonumber\\
      & &\times\exp[{i \over 2\hbar}
                [  x_a^2 M(t_a)(- {\dot{u}(t_a) \over u(t_a)} 
                  +{u(t_b)\dot{v}_s(t_a) \over u(t_a) v_s(t_b)} )       
           +(x_b-x_p(t_b))^2M(t_b){\dot{v}_s(t_b) \over v_s(t_b)}
   \cr&&~~~~~~~~~~~
           -2x_a(x_b-x_p(t_b))M(t_a){\dot{v}_s(t_a) \over v_s(t_b)}
      +2M(t_b)\dot{x}_p(t_b)x_b -2M(t_a)\dot{x}_p(t_a)x_a 
   \cr&&~~~~~~~~~~~
      +2M(t_b)a(t_b)x_b^2-2M(t_a)a(t_a)x_a^2+2b(t_b)x_b-2b(t_a)x_a   
   \cr&&~~~~~~~~~~~
      -M(t_b){\dot{v}_s(t_b) \over v_s(t_b)} x_p^2(t_b)
   \cr&&~~~~~~~~~~~
      -\int_{t_a}^{t_b}(-2f(t)+{M(t)\over v_s^2(t)}
        [x_p(t)\dot{v}_s(t)-\dot{x}_p(t)v_s(t)]^2) dt  ]],
\end{eqnarray}
whose difference from $K^F(b,a)$ is just from the above-mentioned four 
terms and an integral of $f$.
As in the previous sections, one can prove that this kernel does not 
depend on the way of choosing classical solutions.

The n-th order wave function $\psi_n$ can be found from the kernel as;
\begin{eqnarray}
\psi_n^G(x,t)&=& 
     {1\over \sqrt{2^n n!}}({\Omega \over \pi\hbar})^{1\over 4}
     {1\over \sqrt{\rho(t)}}[{u(t)-iv(t) \over \rho(t)}]^{n+{1\over 2}}
\cr&&\times
     \exp[{i\over\hbar}[M(t)a(t)x^2+ (M(t)\dot{x}_p(t)+b(t))x]] 
\cr&&\times
     \exp{[{(x-x_p(t))^2\over 2\hbar}(-{\Omega \over \rho^2(t)}
               +i M(t){\dot{\rho}(t) \over \rho(t)})]}
         H_n(\sqrt{\Omega \over \hbar} {x -x_p(t) \over \rho(t)})
\cr&&\times
         \exp[{i\over\hbar}[-{M(t)\over 2}{\dot{v}(t)\over v(t)} x_p^2(t)
\cr&&~~~~~~~~~~~~~~
            +\int_{t_0}^t (f(z)-{M(z)\over 2}(x_p(z){\dot{v}(z)\over v(z)} 
              -\dot{x}_p(z))^2)dz]].
\end{eqnarray}
Again, $\{u,v\}$ is the set of
arbitrary linear independent homogeneous solutions, and $x_p$ is an
arbitrary particular solution. One can explicitly apply the 
Schr\"{o}dinger equation to this wave functions, to find that they 
indeed satisfy the equation. 

In the Lagrangian of Eq.(1), the conjugate momentum 
$p$ of the coordinate $x$ is written as $p=M\dot{x}+2Max+b$. 
One may interpret $x_p$ as the classical coordinate, and the 
conjugate momentum is then written as 
\be 
p_{p}= M\dot{x}_p+2Max_p+b.
\ee
As in the Sec. II, we define a $\gamma'$ as $\gamma_1'+i\gamma_2'$ 
where
$$ \gamma_1'=\gamma_1 ~~~{\rm and}~~~ 
\gamma_2'=-{M\over\hbar}(2a+{\dot{\rho}\over \rho}).$$
Then, the wave function $\psi_n^G$ can be simply written as
\be
\psi_n^G(x,t) ={1\over \sqrt{2^n n!}}({\gamma_1\over \pi})^{1\over 4}
            e^{i\delta(t)}\exp[-{\gamma'\over 2}(x-x_p)^2 +{i\over \hbar}xp_p]
             H_n(\sqrt{\gamma_1}(x-x_p)),
\ee
where $\delta(t)$ is a real function of $t$. The wave functions agree with 
those in Ref.\cite{Yeon2} except the fact $\delta(t)$ is real which is 
necessary for the conservation of total probability  
$\int \psi_n^{G*}\psi_n^G dx$.
The expression of $\psi_n^G$ in Eq.(39) shows that $\psi_0^G$ is a wave 
function of a Gaussian pure state \cite{Sch}, so the discussions of 
Sec. II are still valid in the general case.

With the wave functions, one can calculate the expectation values of
operators, and the uncertainty relations read; 
\begin{eqnarray}
&_m\!\!<(\Delta x)^2>_m&\!\!_m\!\!<(\Delta p)^2>_m =
(~_m\!\!<x^2>_m -~_m\!\!<x>_m^2)(~_m\!\!<p^2>_m -~_m\!\!<p>_m^2)
\cr
&=&(m+{1\over 2})^2\hbar^2[1+{1\over \Omega^2}
     (2Ma\rho^2+M\rho\dot{\rho})^2],
\\
&_{m+1}\!\!<(\Delta x)^2>_m&\!\!_{m+1}\!\!<(\Delta p)^2>_m\cr 
&=&{1\over\sqrt{2}}
      ({(m+1)\hbar \over \Omega})^{3\over 2}(u+iv)^3
   [{2\sqrt{2\Omega}x_p \over \sqrt{(m+1)\hbar}(u+iv)} -1]
   [2Ma+M{\dot{\rho}\over \rho}+i{\Omega\over \rho^2}]
\cr
&&\times[p_p
  -{1\over 2}\sqrt{(m+1)\hbar\over 2\Omega}
   (u+iv)(2Ma+M{\dot{\rho}\over \rho}+i{\Omega\over \rho^2})],
\\
&_{m+2}\!\!<(\Delta x)^2>_m&\!\!_{m+2}\!\!<(\Delta p)^2>_m\cr
&&=(m+2)(m+1)({\hbar\over 2\Omega})^2
    (u+iv)^4(2Ma+M{\dot{\rho}\over \rho}+i{\Omega\over \rho^2})^2,
\end{eqnarray}
with the notation 
$_n\!\!<O>_m=\int_{-\infty}^\infty \psi_n^{G*} (x,t)O\psi_m^G (x,t)$. 
If we take $u=\rho\cos\theta$ and $v=\rho\sin\theta$, 
then $\Omega=M\rho^2\dot{\theta}$ and the functions 
$\rho(t)$, $\theta(t)$ should satisfy:
$$\ddot{\theta}+2{\dot{\rho}\over\rho} \dot{\theta} 
+{\dot{M}\over M}\dot{\theta} =0,$$
$$\ddot{\rho}+{\dot{M}\over M}\dot{\rho} -\rho\dot{\theta}^2+w^2\rho=0.$$
With these notations, the uncertainty relations in Eq.(40-42) are written 
as;  
\begin{eqnarray}
&_m\!\!<(\Delta x)^2>_m&\!\!_m\!\!<(\Delta p)^2>_m =
       (m+{1\over2})^2\hbar^2[1+{1\over\dot{\theta}^2}
                             (2a+{\dot{\rho}\over\rho})^2],  \\
&_{m+1}\!\!<(\Delta x)^2>_m&\!\!_{m+1}\!\!<(\Delta p)^2>_m \cr
&&=
    {(m+1)^2\over 4}\hbar^2e^{4i\theta}{1\over \dot{\theta}^2}
   [1-{2\sqrt{2M\dot{\theta}}\over\sqrt{(m+1)\hbar}}x_pe^{-i\theta}]
   [2a+ {\dot{\rho}\over\rho} +i\dot{\theta}]\cr
 &&~~\times
     [2a+ {\dot{\rho}\over\rho} +i\dot{\theta}
       -{2\sqrt{2\dot{\theta}}\over\sqrt{(m+1)M\hbar}}
          p_{p} e^{-i\theta}]      \\
&_{m+2}\!\!<(\Delta x)^2>_m&\!\!_{m+2}\!\!<(\Delta p)^2>_m
     ={(m+1)(m+2)\over 4}\hbar^2 e^{4i\theta}{1\over \dot{\theta}^2}
     [2a+ {\dot{\rho}\over\rho} +i\dot{\theta}]^2,
\end{eqnarray}
respectively. The uncertainty relations of Eq.(43,45) 
{\em exactly} agree with those of Ref.\cite{Yeon2}, but the uncertainty 
relation of Eq.(44) differs from the corresponding one there.

The terms which do not affect the classical dynamics of the model 
in Eq.(1) can be written as $L-L^F$. The effects of those
terms on the wave functions could simply be represented by writing
$\psi_n^G$ as;
\be
\psi_n^G(x,t)=\exp[{i\over \hbar}\int^t(L-L^F)(x,{dx\over dz},z) dz]
     \times\psi_n^F(x,t).
\ee 
This relation suggests that $\psi_n^G$ can be obtained from $\psi_n^F$ by 
acting unitary operator $U$;  
\be
U=\exp[{i\over \hbar}(M(t)a(t)x^2+b(t)x +\int^tf(z)dz)].
\ee
By defining operator $O_F, O_G$ as
\be
O_F=-ih {\partial \over\partial t}+H_F, ~~~
O_G=-ih {\partial \over\partial t}+H,
\ee
one may find the relation
\be
UO_FU^\dagger=O_G,
\ee
which proves that the Schr\"{o}dinger equation of general quadratic system
is equivalent to that of the driven
harmonic oscillator through the unitary transformation.

\section{Summary}
The Feynman and Hibbs formulation (or an observation) on the quadratic 
Lagragian system gives a good explanation on the fact that the 
quantum wave function 
can be written in terms of solutions of classical equation of motion.
By developing the observation, we find the kernel and wave functions of 
the general quadratic system in terms of classical solutions. Furthermore,
the kernel is shown to be independent from the choice of classical solutions.
These results are then used to show that the general quadratic system is 
equivalent to the driven harmonic oscillator through a unitary transformation.
This fact \cite{Jackiw} shows that unitary transformation 
(or, canonical transformation, its classical correspondent) could make 
the problem simpler or more complicated,
and could change the uncertainty relations as in Eq.(40-42).
  
\newpage
\section*{Appendix}
\setcounter{section}{1}
\setcounter{equation}{0}
{\def\theequation{%
\Alph{section}.\arabic{equation}}
In this appendix, it will be shown that, if
\be
K(a,b)=\exp[{i\over \hbar}
               (Ax_a^2+Bx_b^2+hx_ax_b+\alpha x_a+\beta x_b +s)]
\ee
where $A,B,h,\alpha$ and $\beta$ are already known functions of 
$t_a,t_b$, then the function $s(t_a,t_b)$ is uniquely determined 
from the initial condition and Schr\"{o}dinger equation.
For the system of Hamiltonian given in Eq.(3), the kernel should 
satisfy the Schr\"{o}dinger equation:
\begin{eqnarray}
i\hbar{\partial \over \partial t_b}K
&=&[{1\over 2M(t_b)}({\hbar\over i}{\partial \over \partial x_b})^2
    -2a(t_b){\hbar\over i}x_b{\partial \over \partial x_b} 
    +{M(t_b)c(t_b)\over 2} x_b^2   \nonumber\\
& & ~~~~~~
    -{b(t_b)\over M(t_b)}{\hbar\over i}{\partial \over \partial x_b}
    +d(t_b)x_b +({b^2(t_b)\over 2M(t_b)}-f(t_b) + i\hbar a(t_b))]K,
\end{eqnarray}
which gives the following differential equations;
\begin{eqnarray}
{\partial A\over \partial t_b}&=&-{h^2 \over 2M(t_b)}\\
{\partial B\over \partial t_b}&=&-{2B^2\over M(t_b)}+4a(t_b)B
                          -{M(t_b)c(t_b)\over 2}\\
{\partial h\over \partial t_b}&=&-{2Bh\over M(t_b)}+2a(t_b)h \\
{\partial \alpha \over \partial t_b}&=&
                     -{h\beta\over M(t_b)}+{b(t_b)\over M(t_b)}h\\
{\partial \beta\over \partial t_b}&=&-{2B\beta\over M(t_b)}+2a(t_b)\beta
                       +2{b(t_b)\over M(t_b)}B -d(t_b)\\
{\partial s\over \partial t_b}&=&-{\hbar\over i}{B\over M(t_b)}
              -{\beta^2\over 2M(t_b)}+{b(t_b)\over M(t_b)}\beta
              -{b^2(t_b)\over 2M(t_b)}+f(t_b)-i\hbar a(t_b).
\end{eqnarray}
With the explicit expressions of $A,B,h,\alpha$ and $\beta$, one may 
check that Eqs.(A.3-7) are satisfied. For example, in the general 
quadratic system considered in section IV, $B$ is given as 
$B(t_a,t_b)={M(t_b)\over 2}{\dot{v}_s(t_b)\over v_s(t_b)}+M(t_b)a(t_b)$
which satisfies Eq.(A.4). 

$s(t_a,t_b)$ can be determined from Eq.(A.8) up to a function 
$g(t_a,t_0)$;
\begin{eqnarray}
s(t_a,t_b)&=&g(t_a,t_0)-{\hbar\over i}\int_{t_0}^{t_b}{B\over M(t)}dt\cr
      & &-\int_{t_a}^{t_b}
        [{\beta^2\over 2M(t)}-{b(t)\over M(t)}\beta
              +{b^2(t)\over 2M(t)}-f(t)+i\hbar a(t)]dt.
\end{eqnarray}
A wave function $\psi$ satisfies the integral equation
\be
\psi(x_b,t_b)=
      \int_{-\infty}^\infty K(x_b,t_b;x_a,t_a)\psi(x_a,t_a)dx_a .
\ee
In the limit of $t_b\rightarrow t_a$, the classical actions approaches
to 
$${M(t_a) \over 2(t_b-t_a)} (x_b-x_a)^2.$$
In order that the relation of Eq.(A.10) be satisfied in the limit, the
kernel should satisfy the relation: 
\be
K(b,a)\rightarrow
  \sqrt{M(t_a) \over 2\pi i\hbar (t_b-t_a)} 
     \exp[{i M(t_a)\over 2\hbar  (t_b-t_a)}(x_a-x_b)^2]
~~~{\rm as}~~~t_b\rightarrow t_a.
\ee
This initial condition determines the $g$ uniquely.

\end{document}